\documentclass [aip,jap,reprint]{revtex4-1}

\usepackage{graphicx}
\begin{document}

\title{Metal-insulator transition in Nd$_{1-x}$Eu$_x$NiO$_3$ probed by specific heat and anelastic measurements}

\author{V. B. Barbeta}
\email{vbarbeta@fei.edu.br}
\affiliation{Departamento de F\'{i}sica, Centro Universit\'{a}rio da
FEI, S\~{a}o Bernardo do Campo, 09850-901, Brazil}

\author{R. F. Jardim}
\affiliation{Instituto de F\'{i}sica,
Universidade de S\~{a}o Paulo, CP 66318, S\~{a}o Paulo, 05315-970, Brazil}

\author{M. S. Torikachvili}
\affiliation{Department of Physics, San Diego State University, San Diego, CA 92182-1233, USA}

\author{M. T. Escote}
\affiliation{Centro de Ci\^{e}ncias Naturais e Humanas, Universidade Federal do ABC, Santo Andr\'{e}, 09210-170, Brazil}

\author{F. Cordero}
\affiliation{Istituto dei Sistemi Complessi, Via dei Taurini 19, 00185, Rome, Italy}

\author{F. M. Pontes}
\affiliation{Departamento de Qu\'{i}mica, Universidade Estadual Paulista, Bauru, 17033-360, Brazil}

\author{F. Trequattrini}
\affiliation{Physics Department, University of Roma La Sapienza, Italy}

\begin{abstract}
Oxides $R$NiO$_3$ ($R =$ rare-earth, $R \neq$ La) exhibit a metal-insulator (MI) transition at 
a temperature $T_{\rm MI}$ and an antiferromagnetic (AF) transition at $T_{\rm N}$.
Specific heat ($C_{\rm P}$) and anelastic spectroscopy measurements were performed in samples 
of Nd$_{1-x}$Eu$_x$NiO$_3$, $0 \leq x \leq 0.35$.
For $x = 0$, a peak in $C_{\rm P}$ is observed upon cooling and warming at 
essentially the same temperature $T_{\rm MI}= T_{\rm N} \sim 195$ K, although
the cooling peak is much smaller. For $x \geq 0.25$, differences between cooling
and warming curves are negligible, and two well defined peaks are clearly observed: one at lower 
temperatures, that define $T_{\rm N}$, and the other one at $T_{\rm MI}$.
An external magnetic field of 9 T had no significant effect 
on these results. The elastic compliance ($s)$ and the reciprocal of the mechanical quality factor
($Q^{-1}$) of NdNiO$_3$, measured upon warming, showed a very sharp peak at essentially 
the same temperature obtained from $C_{\rm P}$, and no peak is observed upon cooling. The elastic modulus 
hardens below $T_{\rm MI}$ much more sharply upon warming, while the cooling and warming curves 
are reproducible above $T_{\rm MI}$. On the other hand, for the sample with $x = 0.35$, 
$s$ and $Q^{-1}$ curves are very similar upon warming and 
cooling. The results presented here give credence to the proposition that the MI 
phase transition changes from first to second order with increasing Eu doping.

\end{abstract}

\maketitle

A number of $R$NiO$_3$ compounds ($R =$ rare-earth, $R \neq$ La) are metallic at high temperatures 
and display a metal-insulator (MI) transition at a temperature $T_{\rm MI}$, which depends
of the $R$ ion-size of the rare-earth. They also exhibit an antiferromagnetic (AF) transition at 
$T_{\rm N}$, due to the spin ordering of the Ni sublattice. For $R =$ Nd and Pr, 
$T_{\rm MI} \approx T_{\rm N}$, while for the other rare-earths $T_{\rm MI}$ is higher than $T_{\rm N}$, 
with $T_{\rm N}$ increasing slightly and $T_{\rm MI}$ decreasing as a function of the $R$ ionic radius.\cite{MED-A}

The magnetic order of NdNiO$_3$ was studied by powder neutron diffraction (PND),
revealing the presence of a wave propagation vector $k = (1/2,0,1/2)$, and an 
unusual up-up-down-down stacking of ferromagnetically (FM) ordered
planes along the simple cubic (111) direction was proposed.\cite {GAR-A}
On the other hand, soft x-ray resonant scattering experiments 
showed that the (1/2,0,1/2) reflection is of magnetic 
origin, without orbital ordering. Besides,
the results were not consistent with 
the spin arrangement proposed by PND, and indicated
a non-collinear antiferromagnetic ordering scheme.\cite{SCA-A}

Recently, high resolution PND experiments in NdNiO$_{3}$ 
unambiguously established the occurrence of two different NiO$_{6}$ octahedra 
at low temperatures, as well as the corresponding change from orthorhombic ($Pbnm$) to 
monoclinic ($P\rm 2_{\rm 1}/n$) symmetry.\cite{GAR-C}
This being the case, a charge ordered state is observed at low temperatures 
and the twofold e$_{g}$ orbital degeneracy is lifted, opening an energy gap. 
Therefore, the low temperature phase may not be classified as
a charge transfer insulator, as originally suggested, but could be better
described as a band insulator.\cite{GAR-C} 

Although much work have addressed the general physical properties 
of these systems, there are still many open questions, regarding the 
role played by the correlation between magnetic and electronic 
properties. Within this context, here we present and discuss measurements of 
specific heat ($C_{\rm P}$) and anelastic spectroscopy near the MI
phase transformation.

Polycrystalline samples of Nd$_{1-x}$Eu$_x$NiO$_3$, $0 \leq x \leq 0.35$, 
were prepared from sol-gel precursors, sintered at temperatures 
$\sim 1000 ^{\rm o}$C, under oxygen pressures up to 80 bar. Details 
of sintering process for preparing these 
samples are described elsewhere. \cite{ESC-A} All samples were 
characterized by X-ray powder diffraction in a 
Brucker D8 Advance diffractometer. The X-ray diffraction patterns showed no extra 
reflections due to impurity phases, and indicated that all samples 
have a high degree of crystallinity.

Specific heat ($C_{\rm P}$) measurements in the temperature range from
2 to 310 K, upon cooling and warming,were performed in a Physical 
Property Measurement System (PPMS) from Quantum Design equipped 
with a superconducting 9 T magnet.

Complex Young's modulus measurements
$E(\omega,T)=E^{\prime} + iE^{\prime\prime}$
were performed as a function of temperature, by electrostatically 
exciting the fundamental flexural modes of the samples, and detecting 
the vibration amplitude. The energy dissipation  
or reciprocal of the mechanical quality factor,
$Q^{-1}(\omega,T) = E^{\prime\prime}/E^{\prime}$, was determined from 
the decay of the free oscillations or from the width of 
the resonance peak. In light of the porosity 
of the sintered materials, the values of elastic compliance 
$s = E^{-1}$ were not absolute, and
therefore were normalized to the $s_{0}$ value, obtained at 
the fundamental frequency $f_{0}=f(T=0)$.

The $C_{\rm P}$ results for NdNiO$_{3}$ are displayed in Figure 1 for both 
the warming and cooling cycles. The sharp peak observed upon warming at 
$T_{\rm MI} = T_{N} \sim 195$ K defines the MI and AF transitions. The 
peak in $C_{\rm P}(T)$ at $T_{\rm MI}$ upon cooling is much reduced. 
The difference in $C_{\rm P}(T)$ near $T_{\rm MI}$ upon cooling
and warming suggests a complex interaction between the crystalline and
magnetic structures, and perhaps that the 
phase transition at $T_{\rm MI}$ has a first order character. Difficulties in
extracting accurate values of $C_{\rm P}(T)$ near first order transitions
using relaxation calorimetry are known. \cite{LAS-A} However, the
large difference between the cooling and warming cycles in this case is
compelling enough to suggest intrinsic behavior.

\begin{figure} [htp]
\centering
\includegraphics [width=0.39\textwidth] {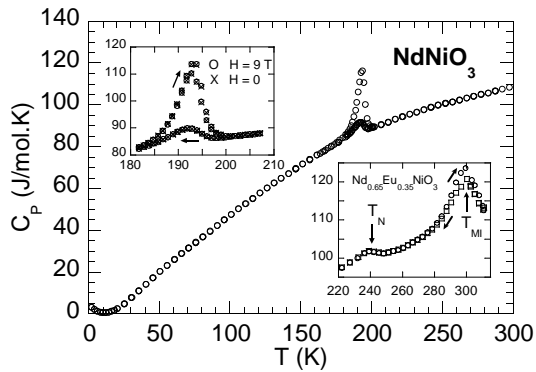}
\caption{\label{fig:epsart1} Temperature dependence of $C_{p}$ for NdNiO$_3$ upon cooling and warming. The upper inset displays the transition region
with $H = 0$ and $H = 9$ T. The lower inset shows the $C_{p}(T)$ data for
the Nd$_{0.65}$Eu$_{0.35}$NiO$_3$ sample.}
\end{figure}

When Nd is partially replaced by Eu, both $T_{\rm MI}$ and $T_{\rm N}$ are shifted to higher temperatures.
For the $x = 0.35$ sample (lower inset of Figure 1), electronic and
magnetic transitions are separated in temperature, and
two peaks in $C_{\rm P}(T)$ are clearly identified.
In this case, there is no significant difference between the
cooling and warming curves. The application of an external 
magnetic field, as high as 9 T, resulted in no appreciable change in $C_{\rm P}(T)$ data, as displayed 
in the upper inset of Figure 1 for the NdNiO$_3$ sample. 
Similar field independent behavior was also observed in the Nd$_{0.65}$Eu$_{0.35}$NiO$_3$ sample (not shown).

\begin{figure} [htp]
\centering
\includegraphics [width=0.39\textwidth] {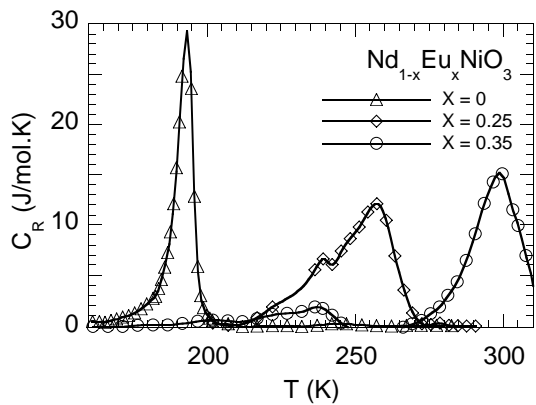}
\caption{\label{fig:epsart2} Temperature dependence
of the specific heat ($C_{R}$) 
of Nd$_{1-x}$Eu$_{x}$NiO$_3$, for three
selected samples, obtained after subtracting
the background contribution (see text). Lines are just
a guide to the eye.}
\end{figure}

A background contribution to $C_{\rm P}(T)$ was subtracted from the curves 
and the resulting specific heat ($C_{\rm R}(T)$) is displayed in Figure 2.
Such a subtraction was performed by excluding the region close to the phase transition in the warming cycle, 
and fitting the resulting curve to a smooth base-line. The resulting curve for the $ x = 0$ sample 
displays a very sharp peak at $T_{\rm MI} = T_{N} \sim 195$ K. However, the partial 
substitution of Nd with Eu results in a separation of the two transitions.
This is clearly seen in the $x = 0.25$ sample which exhibits two peaks: one at $T \sim 240$ K, related 
to the AF transition, and the other one at $T \sim 270$ K, due to the MI transition. 
The separation in temperature of the two transitions is even more
evident in the $x = 0.35$ sample, where the two peaks 
are completely resolved in spite of some smearing in comparison with the $x=0$ sample.

Photoemission spectra (PES) measurements performed across the MI phase 
transition in systems with $T_{\rm MI} = T_{\rm N}$ (NdNiO$_{3}$ and PrNiO$_{3}$),
revealed a temperature-dependent loss of spectral weight near the chemical potential,
extending well below the transition temperature. On the other hand,
samples with $T_{\rm MI} > T_{\rm N}$ (SmNiO$_{3}$ and EuNiO$_{3}$)
showed a quite different behavior. These results were interpreted as
an indication that there is a qualitative difference between 
these two systems, and that there 
is an interplay between magnetic and electronic
properties in samples with $T_{\rm MI} = T_{\rm N}$.\cite{VOB-A}

Similar results were obtained through PES measurements
performed in the metallic phase, 
in samples of Nd$_{1-x}$Sm$_{x}$NiO$_{3}$.\cite {OKA-A}
It was observed that for $x > 0.4$ 
($T_{\rm MI} > T_{\rm N}$) the spectra above $T_{\rm MI}$
shows a pseudogap, different from the spectra 
behavior obtained for $x \leq 0.4$ ($T_{\rm MI} = T_{\rm N}$),
which is typical of a metal. This difference in the nature of the metallic 
state was assumed to be caused by changes in the strength 
of the electronic correlation between the two regions. Another result indicating a crossover in the nature
of the MI phase transition was obtained through magnetic
susceptibility data. Measurements performed in samples of
NdNiO$_{3}$ and Nd$_{0.5}$Sm$_{0.5}$NiO$_{3}$,
after the subtraction of the magnetic rare-earth contribution,
indicated a change from Pauli to Curie-Weiss paramagnetism
with increasing Sm doping in NdNiO$_{3}$ sample.\cite{ZHO-A}
It was also observed that below $T_{N}$
the magnetic susceptibility increases
with decreasing temperature, with no difference 
between the field-cooled and zero-field-cooled
cycles, a situation that is not characteristic of
a system with localized spin. Finally, based on electrical
resistivity measurements, it was suggested that the 
MI phase transition is first order only
when $T_{\rm MI} = T_{\rm N}$, and second order
for $T_{\rm MI} > T_{\rm N}$.\cite{ZHO-A}

\begin{figure} [htp]
\centering
\includegraphics [width=0.39\textwidth] {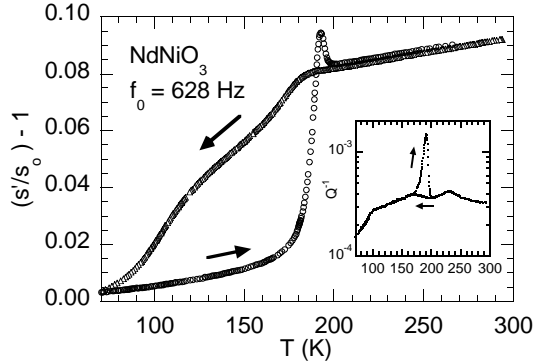}
\caption{\label{fig:epsart3} Elastic compliance $(s/s_{0})-1$
for $x = 0$. The inset shows the corresponding graph of $Q^{-1}$.}
\end{figure}

The elastic compliance curves $(s/s_{0})-1$ obtained for the $x = 0$
sample showed essentially the same 
behavior above $T_{\rm MI} \sim 195$ K upon cooling and
warming, as shown in Figure 3.
Below $T_{\rm MI}$, the hardening is
very gradual upon cooling, extending over more than 100 K.
On the other hand, the softening upon warming is abrupt,
resulting in a very strong hysteresis
between both curves. The energy dissipation (inset
of Figure 3) has a marked peak upon warming at $T \sim 191$ K,
due to the MI phase transition. It was expected to
find such a peak also in the cooling curve, marking
the phase transition, but the presence of this peak
is not evident or completely absent.

The cooling behavior of $(s/s_{0})-1$ below $T_{\rm MI}$ is 
possibly caused by the coexistence of the metallic and insulating 
phases, typical situation of a first-order phase transition.
The coexistence of both phases below $T_{\rm MI}$ has been
reported previously, from electrical resistivity measurements.\cite{GRA-A}
Differently from transport measurements, which mainly reflect
the percolation of the conducting phase, the observed variation of
the elastic compliance below $T_{\rm MI}$ is related to the volumetric
fraction of the transformed phase and indicates that, even at 70 K
(the lowest temperature of these measurements),
some residual metallic phase still remains within the sample.

On the other hand, for the specimen with $x = 0.35$, the elastic
compliance ($s$) and the reciprocal of the mechanical quality factor
($Q^{-1}$) showed no significant hysteresis upon cooling and warming
(not shown). Previous electrical resistivity measurements 
also showed a negligible hysteresis for samples with $x \geq 0.25$,\cite{ESC-B}
indicating that the phase transition above this concentration
is possibly of second order.

In conclusion, we have found that it is possible to distinguish two different regimes in 
the MI phase boundary in Nd$_{1-x}$Eu$_x$NiO$_3$: (i) for
$x < 0.25$, $T_{\rm MI} = T_{\rm N}$, there is a strong 
thermal cycle hysteresis in the $C_{\rm P}(T)$, 
and the transition has a first order character; (ii) for $x \geq 0.25$, $T_{\rm MI} > T_{\rm N}$, the thermal hysteresis is 
negligible, and the MI transition has a second order character. 
These propositions are consistent with features observed in elastic modulus measurements, 
which also showed a strong hysteresis upon cooling and warming for the 
$x = 0$ sample, a feature that has been found to be insignificant in the $x = 0.35$ sample.
Previous magnetic susceptibility and electrical
resistivity measurements in a series of Nd$_{1-x}$Eu$_x$NiO$_3$ samples are consistent with
this picture.\cite{ESC-B}

\begin{acknowledgments}

This work was supported by the Brazilian agency
Funda\c{c}\~{a}o de Amparo \`{a} Pesquisa do Estado de S\~{a}o
Paulo (FAPESP) under Grant No. 2005/53241-9. Three of us (R.F.J., M.T.E., and F.M.P.)
acknowledge the Conselho Nacional de Desenvolvimento
Cient\'{i}fico e Tecnol\'{o}gico (CNPq) for fellowships. M.S.T. gratefully acknowledges
support of the National Science Foundation under Grant No. DMR-0805335

\end{acknowledgments}

\end{document}